Doppler shift experiments with source in periodic motion:
Parametrized Doppler shift formulas.

Bernhard Rothenstein, Physics Department, University of Timisoara, 1900, Timisoara, Romania
Albert Rothenstein, Centre for Vision Research, York University, Toronto, Canada

*Abstract.*
*Doppler shift formulas are derived for two less studied scenarios: stationary receiver and source in harmonic oscillatory motion and stationary receiver and source in uniform circular motion. For each of the scenarios we derive a formula, one which works when the emission period is small enough that it can be considered that two successive signals are emitted from the same point in space (locality assumption) and another which takes into account that two successive signals are emitted from two different points in space (non-locality assumption). The results furnished by the two Doppler shift formulas are compared, showing that increasing the emission frequency decreases the difference between the results obtained with the two formulas.*

1. Introduction

Doppler shift experiments are performed in college laboratories. The experimental devices used and the results obtained are presented by many authors. Different scenarios are followed, from the simplest longitudinal Doppler effect (stationary observer and source moving with constant speed along the line which joins them) [1] to the more sophisticated ones, involving an accelerating source [2], or source in uniform motion but with an oblique angle of incidence to the received wave crests [3].
The experimental results are interpreted using a Doppler shift formula which holds only in the case of the plane wave approximation (very big distance between receiver and source) or in the case of the very small period approximation [3], even if it is obvious that the scenarios followed do not meet the imposed conditions. In the case of the very small period approximation, the periods are small enough that it can be considered that the moving source emits two successive wave-crests from the same point in space. This idealized situation is called locality. The simple fact that the source moves is associated with the fact that it emits two successive wave-crests from two different points in space and we call this situation non-locality, [4].
We present first a derivation of the Doppler shift formula in the locality assumption. Receiver R is located at the origin O of its rest frame K(XOY). The position of the source S in K(XOY) is defined by its polar coordinates $S(r,\theta_S)$, (Figure 1). S emits an acoustic signal (wave-crest) at a time $t_S$, which is received by R at a time $t_R$. The two times are related by

$$t_R = t_S + \frac{r}{U} \qquad (1)$$

Differentiating both sides of Equation (1) we obtain

$$dt_R = dt_S + \frac{dr}{U} \quad (2)$$

Taking into account that, by definition,

$$\frac{dr}{dt_S} = v\cos\theta \quad (3)$$

represents the radial component of the instantaneous velocity v of the source, Equation (3) leads to the result

$$dt_R = dt_S\left(1 + \frac{v}{U}\cos\theta\right) \quad (4)$$

It is obvious that Equation (4) holds only in the case of the very small period approximation, when we can consider that $dt_S$ and $dt_R$ represent the very small periods at which the wave-crests are emitted and received, respectively. Introducing the concept of "very high frequency," Equation (4) is presented in many cases as

$$D = \frac{f_R}{f_S} = \frac{1}{1 + \frac{v}{U}\cos\theta} \quad (5)$$

where D represents a Doppler factor and where $f_R=1/dt_R$ and $f_S=1/dt_S$ represent the frequencies at which the signals are emitted and received, respectively.

In the non-locality approach, we consider that S emits two successive wave-crests (signals) at $t_S$ and $t_S+T_S$, where $T_S$ represents the constant period at which the signals are emitted. The emissions take place from the positions of the source $S_1(r_1,\theta_{S,1})$ and $S_2(r_2,\theta_{S,2})$. R(0,0) receives the two signals at the times

$$t_{R,1} = t_S + \frac{r_1}{U} \quad (6)$$

and

$$t_{R,2} = t_S + T_S + \frac{r_2}{U} \quad (7)$$

so the time interval between the reception of the two signals is given by

$$T_{R,1,2} = t_{R,2} - t_{R,1} = T_S + \frac{r_2}{r_1} \quad (8)$$

the result depending on the law which defines the motion of the source relative to the receiver.

The purpose of our paper is to follow scenarios which ensure that during the emission of two successive signals the velocity of the source and the direction along which the signals are emitted can change. In each case, locality and non-locality approaches are presented and the obtained results are compared.

2. Doppler shift formula for stationary observer and source in harmonic motion.

Observer R is located at the origin O of the K(XOY) reference frame and the source performs an oscillatory harmonic motion along the OY axis (Figure 2). We have during a period of the oscillatory motion, longitudinal Doppler effects with outgoing source and increasing velocities, outgoing source with decreasing velocity, incoming source with increasing velocity and incoming source with decreasing velocity, if at t=0 the source S is located in front of the receiver, in a state of rest. All this peculiarities of the oscillatory motion influence the Doppler shift in a characteristic way, increasing or decreasing it. As we see from Figure 2, at a time t the source is at a distance

$$y = A(1 - \cos\omega t_S) \tag{9}$$

from the source, $\omega$ representing the circular frequency and A the amplitude. If a signal is emitted at a time $t_S$, it is received by R at a time $t_R$ given by

$$t_R = t_S + \frac{A}{U}(1 - \cos\omega t_S) \tag{10}$$

Differentiating both sides of Equation (10), we obtain the following Doppler shift formula

$$dt_R = dt_S\left(1 + \frac{V}{U}\sin\omega t_S\right) \tag{11}$$

Expressed as a function of frequencies, Equation (11) becomes

$$D = \frac{f_R}{f_S} = \frac{1}{1 + \frac{V}{U}\sin\omega t_S} \tag{12}$$

where $V = A\omega$ represents the amplitude of the velocity of the source and $v = A\omega\sin\omega t_S$ represents its instantaneous velocity.

In the non-locality approach, we consider that the successively emitted signals are labeled 0,1,2…n…N, where N represents the total number of signals emitted during a period of oscillation $T = 2\pi/\omega$. If $T_S$ represents the constant period at which the signals are emitted, it is related to T by

$$NT_S = T = \frac{2\pi}{\omega} \tag{13}$$

The n-th emitted wave-crest is received by R at time

$$t_{R,n} = nT_S + \frac{A}{U}(1 - \cos\omega n T_S) \tag{14}$$

so the time interval between the reception of the (n-1)-th and the n-th emitted signal is given by

$$T_{R,n-1,n} = t_{R,n} - t_{R,n-1} = T_S + \frac{A}{U}\left(\cos 2\pi \frac{n-1}{N} - \cos 2\pi \frac{n}{N}\right) \tag{15}$$

Equation (15) leads to the Doppler shift formula

$$T_{R,n-1,n} = T_S + \frac{AN\omega}{2\pi U}\left(\cos 2\pi \frac{n-1}{N} - \cos 2\pi \frac{n}{N}\right) \tag{16}$$

Expressed as a function of frequencies, Equation (16) becomes

$$D_{n-1,n} = \frac{f_{R,n-1,n}}{f_S} = \left(1 + \frac{VN}{\pi U}\sin\frac{\pi}{N}\sin\frac{\pi(2n-1)}{N}\right)^{-1} \tag{17}$$

In order to compare the results furnished by Equations (12) and (17), we present the first one as

$$D = \left(1 + \frac{V}{U}\sin 2\pi \frac{n}{N}\right)^{-1} \tag{18}$$

allowing for a continuous variation of n.

The experimental device allows changing N and V/U. Increasing N we reduce the time interval between the emission of two successive signals, a fact which favors the locality assumption. Therefore, the difference between the results furnished by Equations (17) and (18) decreases. This fact is illustrated by Figure 3 where we present the variation of D and $D_{n-1,n}$ with n for N=6, N=12 and N=100 for the same value of v/U=0.5. Increasing v/U we increase the time interval between the emissions of two successive signals, increasing the difference between the results furnished by the two available formulas.

3. Doppler shift formula for stationary observer and source in uniform circular motion.

The scenario under consideration involves a stationary observer R and a source S performing a uniform circular motion (Figure 4). It has high pedagogical potential because during a period of rotation the following typical scenarios are reproduced: longitudinal Doppler effect with outgoing source, Doppler effect with an outgoing source and oblique incidence of the signals, transversal Doppler effect, Doppler effect with incoming source and oblique incidence and, finally, longitudinal Doppler effect with incoming source.

The relative positions of R and S are presented in Figure 4, at time $t_S$, considering that at $t_S=0$, S is located in front of R. At time $t_S$ the distance between R and S is given by

$$RS = 2A\sin\frac{\omega t_S}{2} \qquad (19)$$

where A represents the radius of the circle and $\omega$ the circular frequency.

Consider that at time $t_S$, S sends an acoustic signal which is received by R at time $t_R$, the two times are related by

$$t_R = t_S + \frac{2A}{U}\sin\frac{\omega t_S}{2} \qquad (20)$$

Differentiating both sides of Equation (20) we obtain the Doppler shift formula

$$\frac{dt_R}{dt_S} = 1 + \frac{A\omega}{U}\cos\frac{\omega t_S}{2} \qquad (21)$$

which relates the "very small" periods of emission and reception, respectively. Expressed as a function of the "very big" emission and reception frequencies, Equation (21) leads to the following characteristic Doppler factor

$$D = \frac{f_R}{f_S} = \left(1 + \frac{v}{U}\cos\frac{\omega t_S}{2}\right)^{-1} \qquad (22)$$

where $v=A\omega$ represents the constant magnitude of the velocity of the source.

Taking into account the non-locality in the emission of the signals, we consider that S emits the n-th signal at time $t_S=nT_S$ which is received by R at a time $t_{R,n}$ given by

$$t_{R,n} = nT_S + \frac{2A}{U}\sin\frac{\omega nT_S}{2} \qquad (23)$$

The time interval between the reception of the (n-1)-th and the n-th signal is given by

$$T_{R,n-1,n} = t_{R,n} - t_{R,n-1} = T_S + \frac{2A}{U}\left(\sin\frac{\omega nT_S}{2} - \sin\frac{\omega(n-1)T_S}{2}\right) \qquad (24)$$

Equation (24) leads to the following Doppler shift formula

$$\frac{T_{R,n-1,n}}{T_S} = 1 + \frac{2A}{UT_S}\left(\sin\frac{\omega n T_S}{2} - \sin\frac{\omega(n-1)T_S}{2}\right) \tag{25}$$

Let N be the total number of signals emitted during a period of rotation $T=2\pi/\omega$. Obviously, we have

$$NT_S\omega = 2\pi \tag{26}$$

Introducing the finite frequencies $f_{R,n-1,n}=(T_{R,n-1,n})^{-1}$ and $f_S=(T_S)^{-1}$, we can define the characteristic Doppler factor

$$D_{n-1,n} = \frac{f_{R,n-1,n}}{f_S} = \left(1 + \frac{Nv}{\pi U}\left(\sin\frac{\pi n}{N} - \sin\frac{\pi(n-1)}{N}\right)\right)^{-1} \tag{27}$$

Taking into account Equation (26), we can present Equation (22) as

$$D = \left(1 + \frac{v}{U}\cos\frac{\pi n}{N}\right)^{-1} \tag{28}$$

where we allow a continuous variation for n.

In order to compare the results obtained working with Equation (27) (non-locality assumption) or with Equation (28) (locality assumption), we present in Figure 5 the variation of D and $D_{n-1,n}$ with n for the same value v/U=0.5 and three different values of N (6,12,100). As expected, increasing N decreases the difference between the results furnished by the two Equations, as high values of N favor locality.

4. Conclusions

The study of two different Doppler shift experiments involving a stationary receiver and source in periodic motion (harmonic oscillatory motion and uniform circular motion) shows that we can derive a Doppler shift formula in the "very small period assumption" (two successive signals are emitted from the same point in space and a Doppler shift formula which takes into account the fact that two successive signals are emitted from two different points in space). In the latter case, the Doppler shift formulas are parameterized, being expressed as a function of the order number of the emitted signals. Increasing the emission frequency diminishes the difference between the results furnished by the two formulas.

We consider that it is necessary to make a net distinction between the locality and the non-locality assumptions, not only from a pedagogical, but also from a practical point of view.

References

[1] Bernhard Rothenstein, "Teaching the Doppler shift using a machine gun analogy," Phys.Teach. 39, pp. 468-469 (2001)
[2] T.J.Benski,S.E.Frey, "Computer sound card assisted measurement of the acoustic Doppler effect for accelerated and unaccelerated sound sources," Am.J.Phys.69, pp. 1231-1235 (2001)
[3] Daniel R. Frankl, "General treatment of the Doppler effect in special relativity," Am.J.Phys. 52, pp. 374-375 (1984)

[4] William Moreau, "Non-locality in frequency measurement of uniformly accelerated observer," Am.J.Phys.60, pp. 561-564 (1992)

Captions

Figure1. Scenario for deriving the Doppler shift formula in the locality assumption (source emits two successive wave crests from the same point in space).

Figure2. Scenario for deriving the Doppler shift formula in the case of a stationary receiver and source in harmonic oscillatory motion.

Figure3. The variation of the Doppler factors D (locality assumption) and $D_{n-1,n}$ (non-locality assumption) with the order number n of the emitted wave-crest for the same value of V/U=0.5 and different values of the total number N of signals emitted during a period of the oscillatory motion. Two successive periods of oscillations are presented.
Figure 3a N=6
Figure 3b N=12
Figure 3c N=100

Figure4. Scenario for deriving the Doppler shift formula in the case of a stationary receiver and source in uniform circular motion.

Figure5. The variation of the Doppler factors D (locality assumption) and $D_{n-1,n}$ (non-locality assumption) with the order number n of the emitted wave-crest for the same value of v/U=0.5 and different values of the total number N of signals emitted during a period of the circular motion. Two successive periods of oscillation are presented.
Figure 5a N=6
Figure 5b N=12
Figure 5c N=100

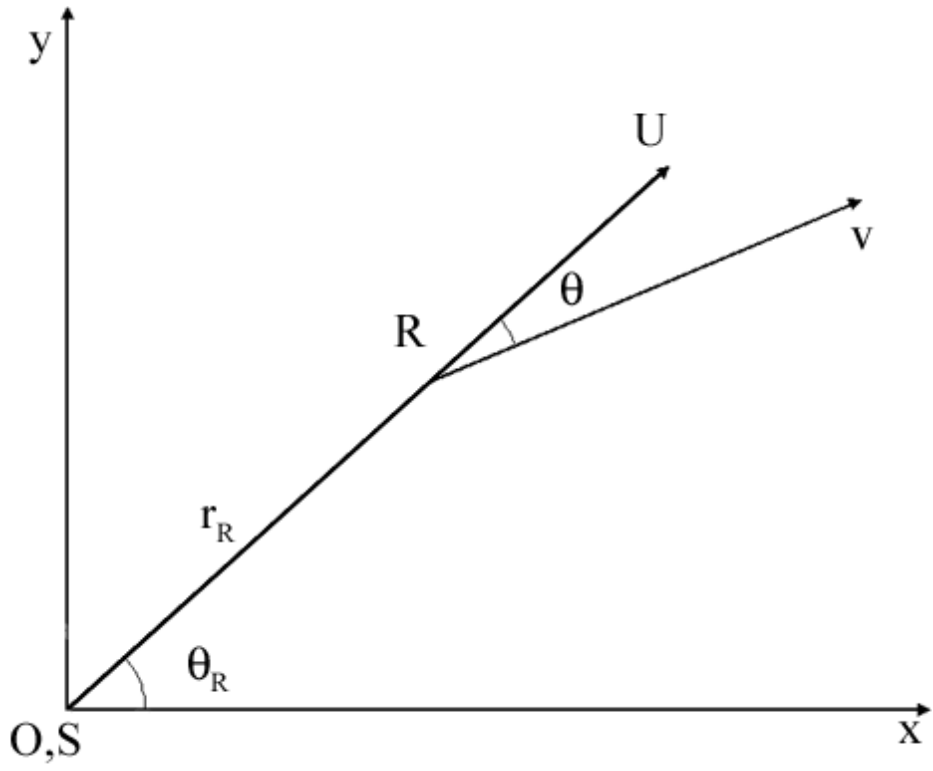

Fig. 1

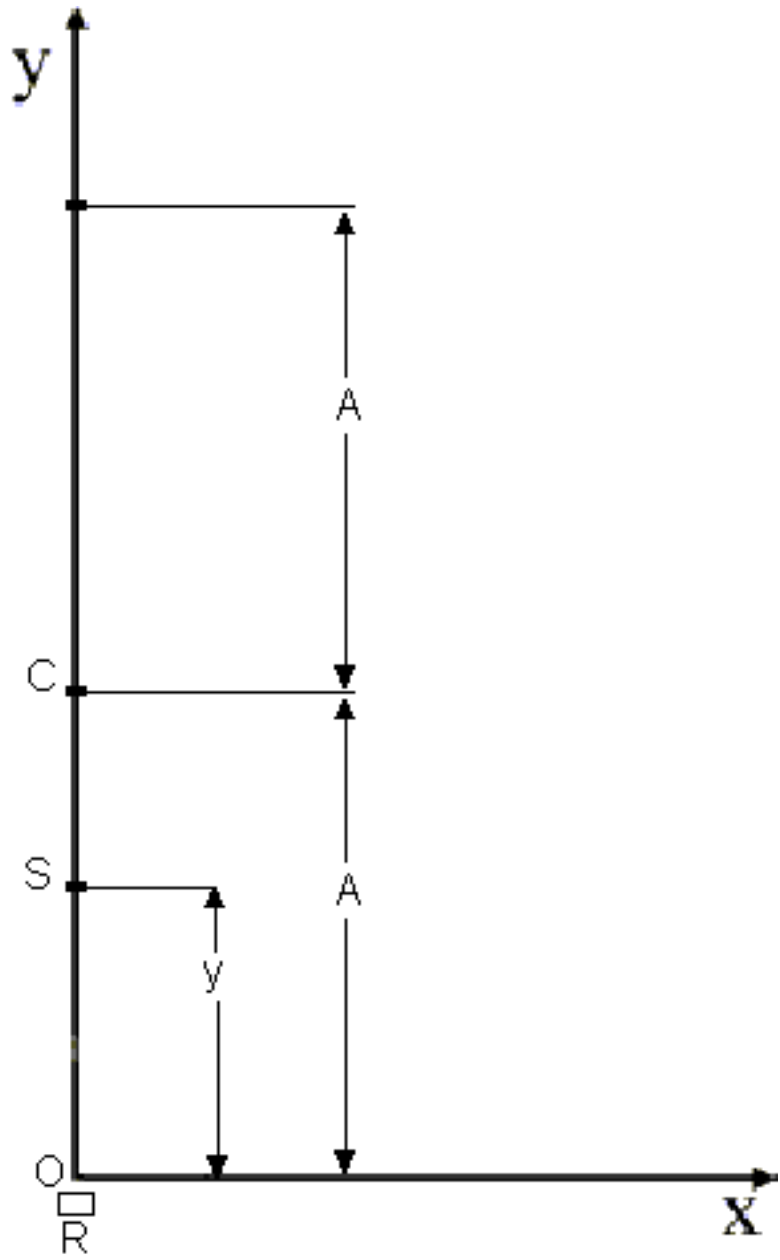

Fig. 2

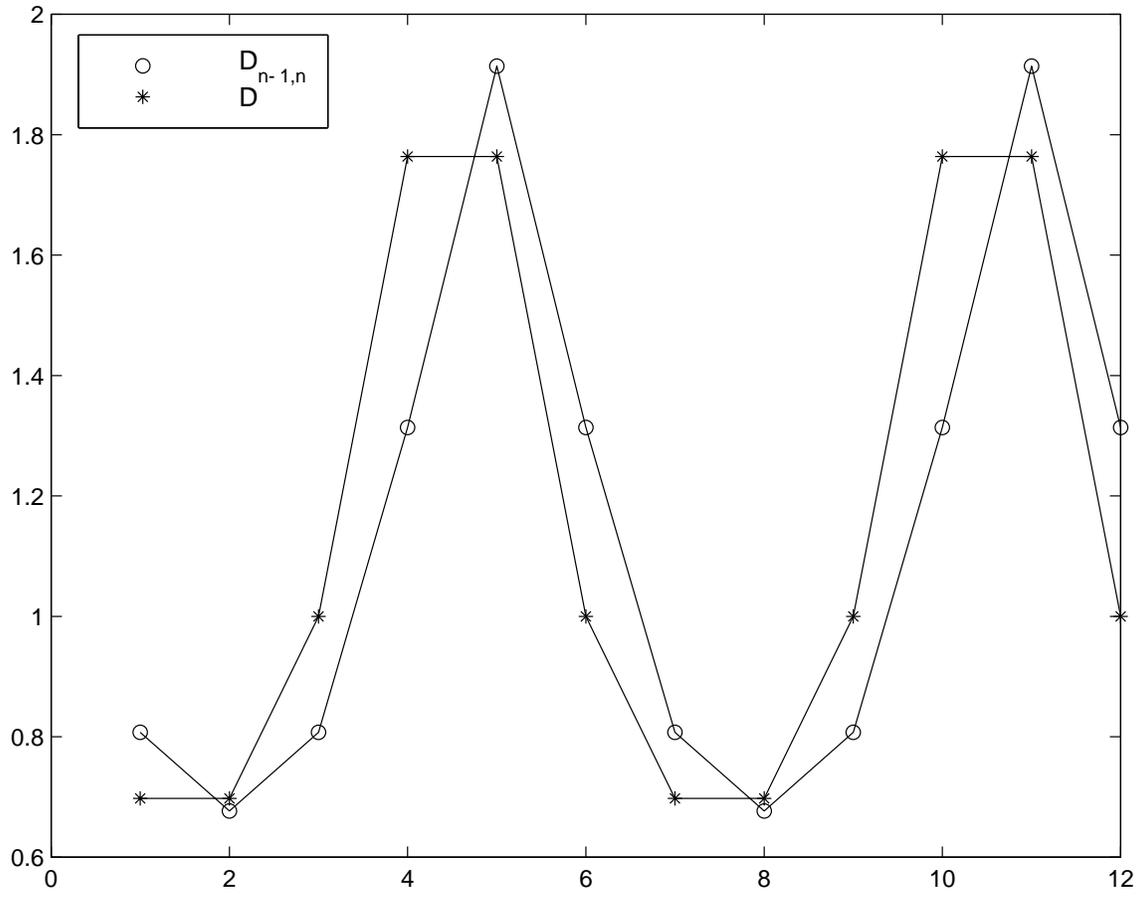

Fig. 3a

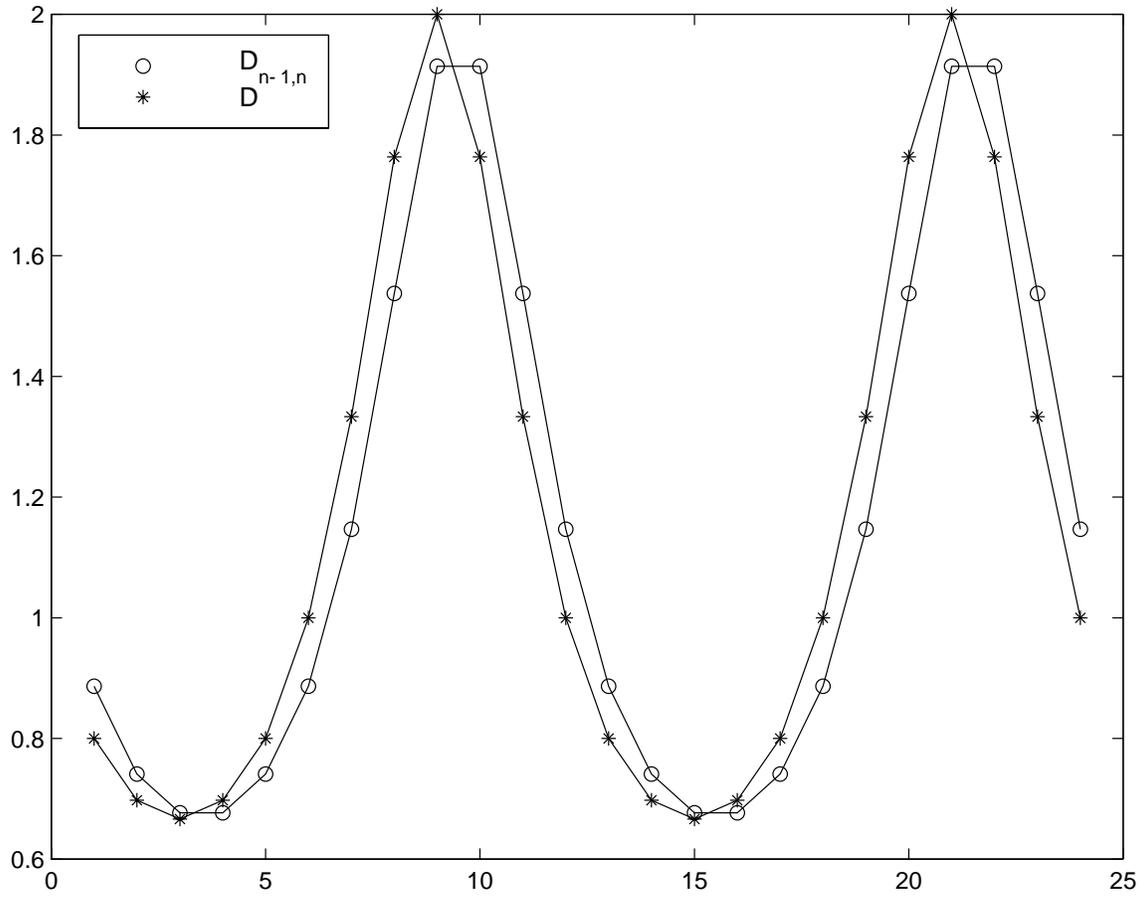

Fig. 3b

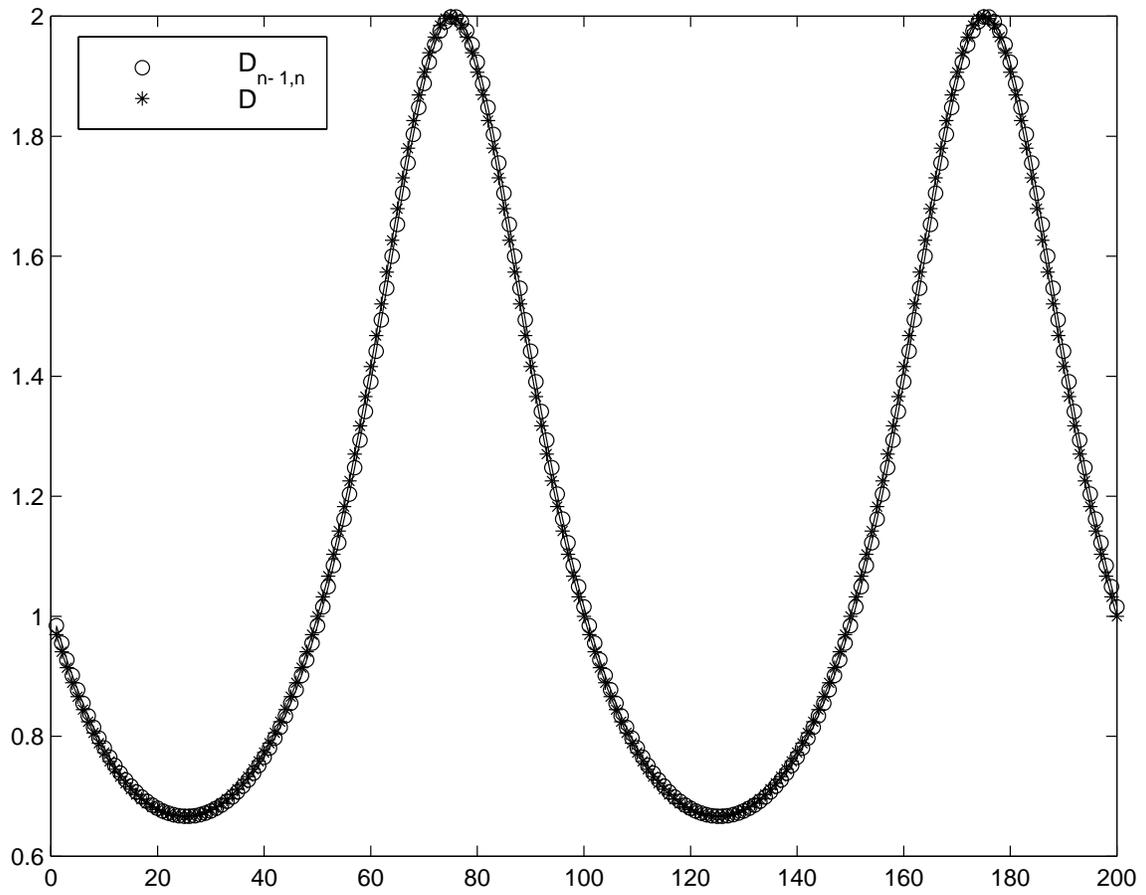
Fig. 3c

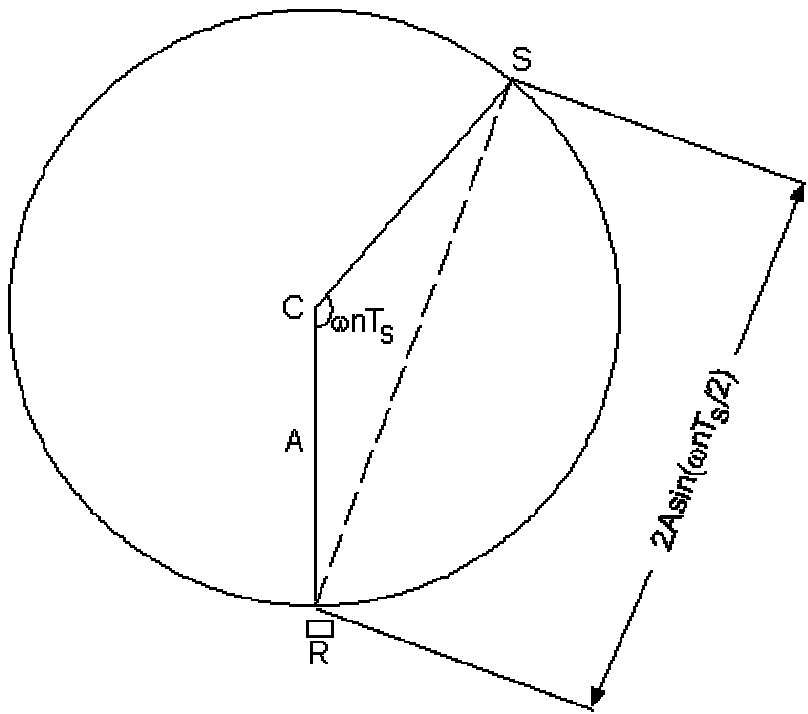

Fig. 4

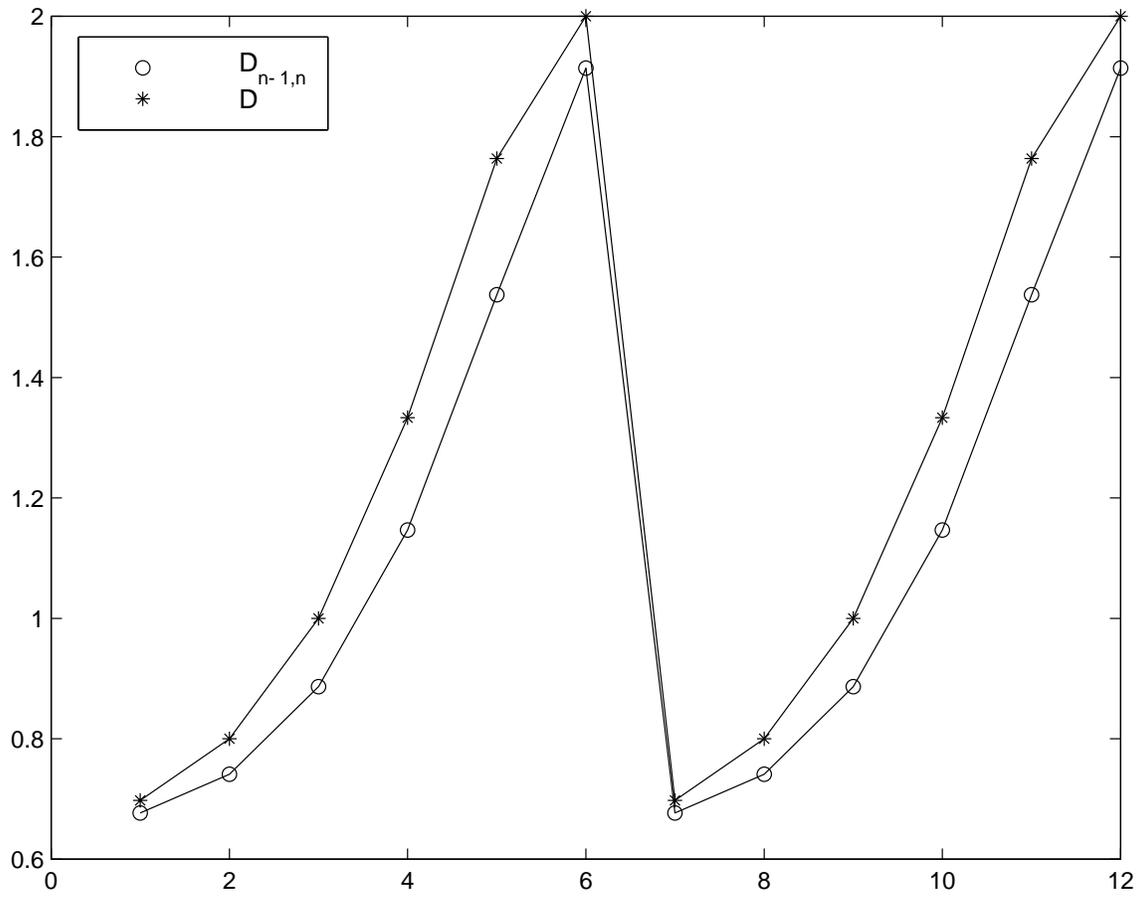

Fig. 5a

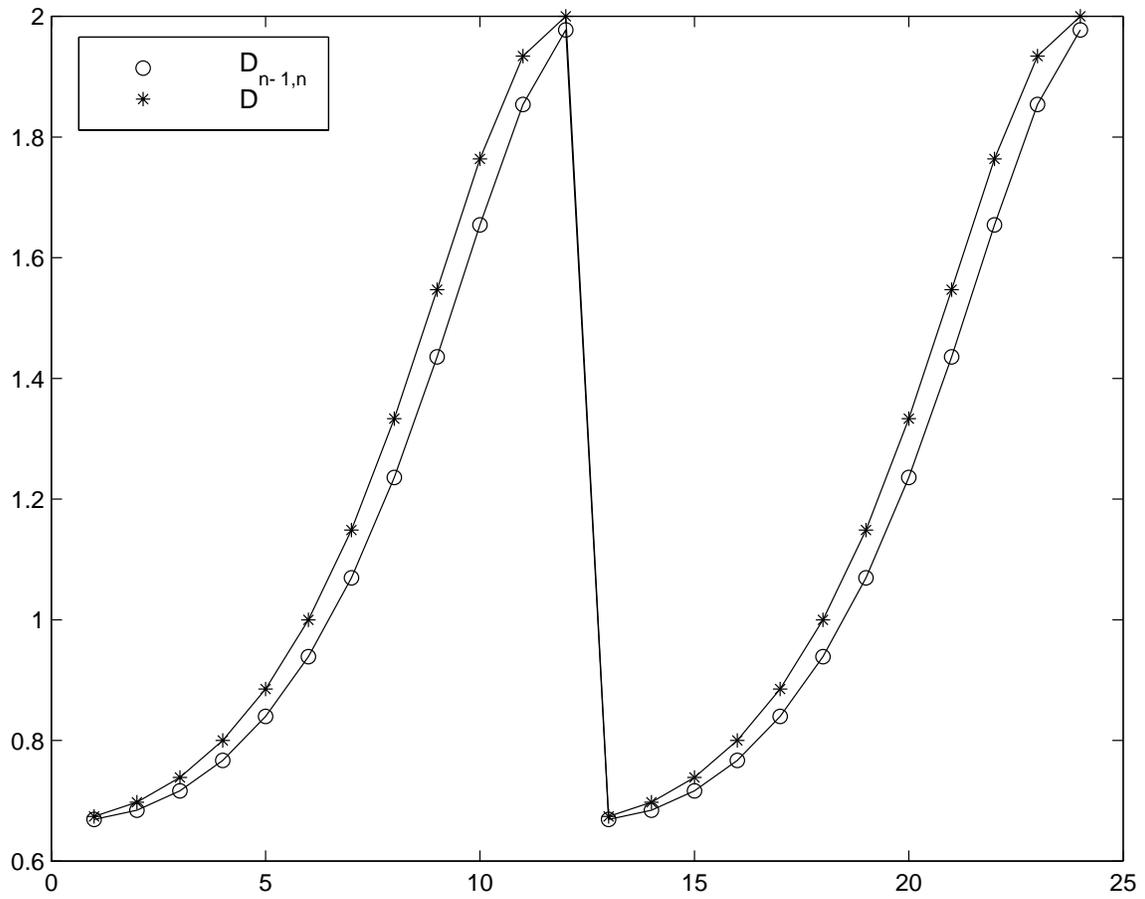

Fig. 5b

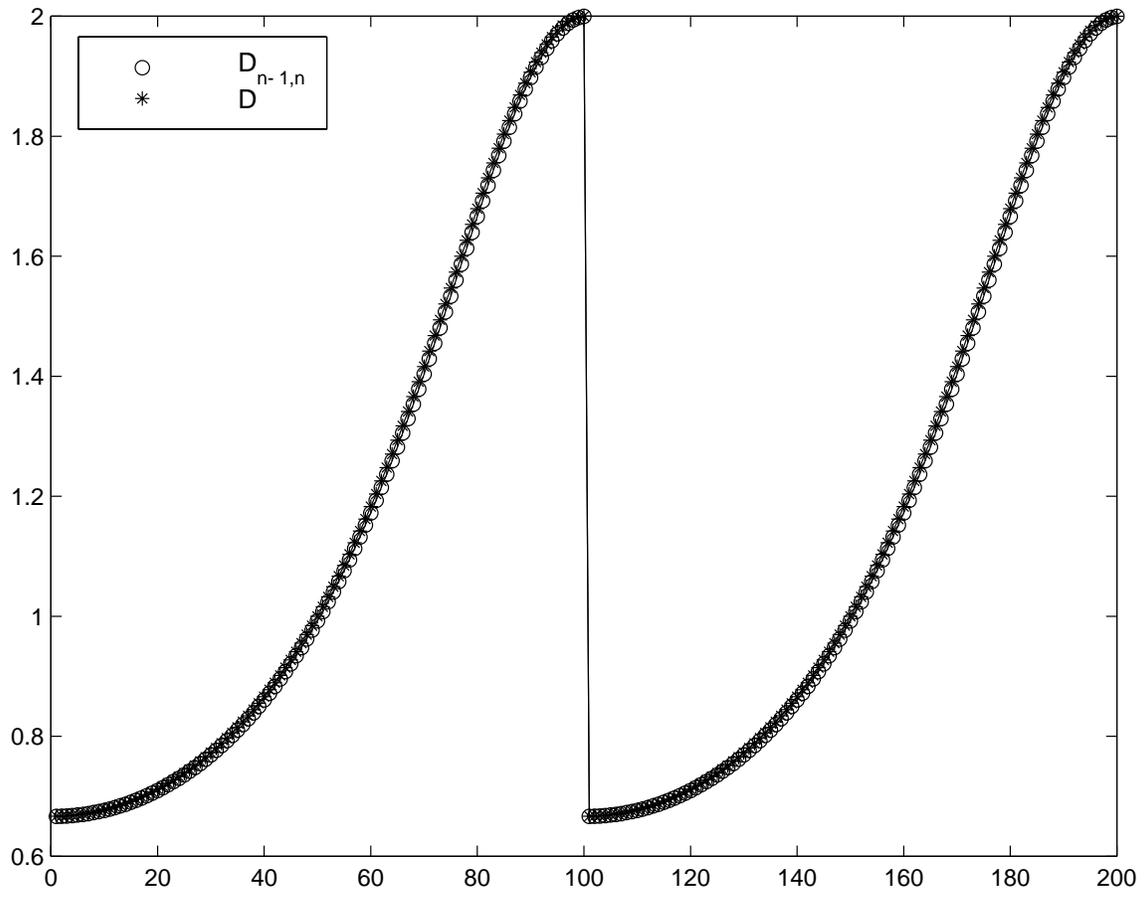

Fig. 5c